\documentstyle[12pt]{article}

\def\rf#1{(\ref{eq:#1})}
\def\lab#1{\label{eq:#1}}
\def\nonu{\nonumber}
\def\br{\begin{eqnarray}}
\def\er{\end{eqnarray}}
\def\be{\begin{equation}}
\def\ee{\end{equation}}

\def\({\left(}
\def\){\right)}

\newcommand{\ct}[1]{\cite{#1}}
\newcommand{\bi}[1]{\bibitem{#1}}
\def\lskip{\vskip\baselineskip\vskip-\parskip\noindent}
\relax

\def\tr{\mathop{\rm tr}}

\newcommand{\sbr}[2]{\left\lbrack\,{#1}\, ,\,{#2}\,\right\rbrack}

\def\b{\beta}

\def\d{\delta}

\def\h{{1\over 2}}
\def\l{\lambda}

\def\o{\over}

\def\pa{\partial}
\def\pr{\prime}

\def\ra{\rightarrow}
\def\s{\sigma}

\def\tp0{\Theta_{+}^{(0)}}
\def\tm0{\Theta_{-}^{(0)}}

\def\u2{\mid u\mid^2}
\def\z2{\mid z\mid^2}

\def\vecgrad{{\vec{\nabla}}}

\def\cj{{\cal J}}

\def\one{\hbox{{1}\kern-.25em\hbox{l}}}
\def\0#1{\relax\ifmmode\mathaccent''7017{#1}%
        \else\accent23#1\relax\fi}

\def\PRL#1#2#3{{\sl Phys. Rev. Lett.} {\bf#1} (#2) #3}
\def\NPB#1#2#3{{\sl Nucl. Phys.} {\bf B#1} (#2) #3}

\def\PRD#1#2#3{{\sl Phys. Rev.} {\bf D#1} (#2) #3}

\def\PLB#1#2#3{{\sl Phys. Lett.} {\bf #1B} (#2) #3}

\def\PR#1#2#3{{\sl Phys. Reports} {\bf #1} (#2) #3}

\def\LMP#1#2#3{{\sl Letters in Math. Phys.} {\bf #1} (#2) #3}

\def\MPLA#1#2#3{{\sl Mod. Phys. Lett.} {\bf A#1} (#2) #3}

\begin{document}

\begin{titlepage}
\vspace*{-2 cm}
\noindent
December, 2000 \hfill{IFT-P.085/2000}\\
hep-th/0010168 \hfill{}

\vskip 3cm

\begin{center}
{\Large\bf  Infinite symmetries in the   Skyrme model}
\vglue 1  true cm
L.A. Ferreira$^1$ and J. S\'anchez Guill\'en$^2$\\

\vspace{1 cm}

$^1${\footnotesize Instituto de F\'\i sica Te\'orica - IFT/UNESP\\
Rua Pamplona 145\\
01405-900, S\~ao Paulo - SP, Brazil}\\

\vspace{1 cm}

$^2${\footnotesize Departamento de F\'\i sica de Part\'\i culas,\\
Facultad de F\'\i sica\\
Universidad de Santiago\\
E-15706 Santiago de Compostela, Spain}

\medskip
\end{center}

\normalsize
\vskip 0.2cm

\begin{abstract}
We show that the Skyrme theory possesses a
submodel with an infinite num\-ber of local conserved currents. 
The constraints leading to the submodel explore a decomposition of $SU(2)$ 
with a complex field parametrizing the symmetric space $SU(2)/U(1)$ and a real
field in the direction of $U(1)$. 
We demonstrate that the Skyrmions
of topological charges $\pm 1$  belong to such integrable sector
of the theory.  
Our results open  ways to the development of exact 
methods, compensating for the non-existence of a BPS type sector in the 
Skyrme theory.

\end{abstract}

\vskip 2cm


\end{titlepage}

Skyrme introduced in the sixties a very original idea which nowadays is
widely used in Physics and Mathematics, that the classical soliton solutions
of a 
given field theory may correspond to particles in its quantum spectrum. He
proposed a theory, now called Skyrme model \ct{skyrme}, to describe the low
energy interactions of hadrons, where the nucleons are interpreted as
topological solitons. The model found applications not only in
Nuclear Physics, but also in Particle and Condensed Matter Physics
\ct{makha,brown,gisiger}. The theory received a boost in the eighties when it
was shown to correspond indeed to the low energy limit of QCD for large $N_c$
\ct{witten,bala}. The soliton solutions have been calculated by numerical
methods \ct{braaten,sut,manton} and there is now a vast amount of evidence
that Skyrme's first spherical Ansatz (the hedgehog) does lead to the solutions
of unit topological charge. The  generalizations of that Ansatz  based on
rational maps lead to good approximations of the minimal energy solutions
\ct{sut,manton}. Such Ans\"{a}tze are inspired in the BPS monopoles by sharing 
its symmetries and  instanton methods. But unlike the well understood
instantons, one easily shows that in three space  dimensions skyrmions cannot
saturate self-duality bounds. The non-existence of a BPS type sector is perhaps
one of the factors  preventing the development of exact methods for the Skyrme
model. In this letter we propose a way of addressing such problems by showing
that there exist a submodel in the Skyrme theory possessing an infinite number
of local conserved currents. They form in fact a continuous set of currents
since they are parametrized by  special functionals of the fields. The submodel
is constructed by imposing Lorentz 
invariant constraints following the methods of \ct{afsg} for studying
integrable theories in any dimension. The constraints explore a suitable
decomposition of the $SU(2)$ group elements where the Skyrme's fields are a
complex scalar field $u$ parametrizing the symmetric space $SU(2)/U(1)$ and a
real scalar field $\zeta$ parametrizing the $U(1)$ subgroup. It turns out that,
in the static case,  particular solutions to the constraints are obtained 
by taking $\zeta$ to depend only upon the radial variable $r$, and $u$ to be a
meromorphic function of a complex variable $z$, built out of the two angles of
the spherical polar coordinates. 
The original Skyrme hedgehog  Ans\"{a}tze for topological charges $\pm 1$ fall 
into such class of configurations and so solve our constraints. We proof
that they solve indeed our submodel equations of motion. The candidates for
higher 
charges solutions were constructed  numerically in \ct{sut}, and it would be
very interesting to verify if they satisfy our constraint. We point out that
the existence of such submodel containing the Skyrme's hedgehog and possessing
an infinite number of conservation laws is highly
non-trivial. It became possible now, about forty years after Skyrme's
original paper, due to the recent developments on the study of integrable
theories in higher dimensions \ct{afsg}.  

Another  interesting point is that the generalizations to higher charges of the
original Skyrme's hedgehog obtained through the rational
map Ans\"{a}tze are particular solutions of our constraints. We show 
that such configurations can not be solutions of Skyrme's model. However,
since they provide approximations to the numerical candidates to true
solutions \ct{manton},  our constraints may give hints on how to correct them
to  
obtain 
of a new Ans\"{a}tze for higher charges solutions. 

In any case, our results  
establish an interesting connection between soliton solutions and infinite
number of conservation laws in $3+1$ field theory of great interest in
Physics. Similar
results have already been found for the Skyrme-Faddeev model \ct{sf} and other
related $(3+1)$-dimensional theories defined on $S^2$ \ct{nice}, as well as in
theories defined on homogenous spaces like $CP^N$ and chiral models
\ct{gianzo,erica,trends,susuki}. 

The Lagrangian density for the Skyrme model, with the pion mass set to zero,
can be written as \ct{skyrme,makha,brown}\footnote{We are using the
space-time metric $ds^2= \( dt\)^2 - \( d {\vec x}\)^2$, and so the relative
signs in the Lagrangian are set by the positivity of the energy.} : 
\be
L =  {f_\pi^2 \over 4}
\tr \left ( U^\dagger \pa_{\mu} U U^\dagger \pa^{\mu} U \right ) -
{1 \over 32 e^2}
\tr \left [ U^\dagger\pa_{\mu} U,U^\dagger\pa_{\nu} U \right ]^2
\ee
where $f_\pi$ and $e$ are phenomenological constants, and $U$ is a $SU(2)$
unitary matrix.  
The equations of motion  derived from this Lagrangian are:
\be
\pa^\mu J_{\mu} = 0 \; ; \qquad {\rm with} \qquad  
J_{\mu} \equiv U^\dagger \pa_\mu U - \l 
\sbr{U^\dagger \pa^{\nu}U}{\sbr{U^\dagger \pa_{\mu} U}{U^\dagger \pa_{\nu}U}}
\lab{skyrmeeqom}
\ee
where $\l=\frac{1}{4 e^2 f_\pi^2}$.

We shall use a special parametrization of the fields of Skyrme model which
leads quite naturally to a submodel possessing an infinite number of
conservation laws and containing the unit charge soliton solutions. 
Denote $U \equiv e^{i \zeta_j \tau_j}$, 
where $\tau_j$, $j=1,2,3$, are the Pauli matrices. Then one gets that
\be
U = e^{i \zeta T} = \one \; \cos \zeta + i \; T \; \sin \zeta 
\ee
where $\zeta \equiv \sqrt{\zeta_1^2 + \zeta_2^2 + \zeta_3^2}$, and
\br
T \equiv \frac{1}{1+\u2}\; \( 
\begin{array}{cc}
\u2 - 1 & -2 i u\\
2 i u^* & 1 - \u2
\end{array}\) 
\er
The complex field $u$ is obtained by making the stereographic projection of
the unit vector $\zeta_i/\zeta$, as
\be
\frac{{\vec \zeta}}{\zeta} \equiv 
\frac{1}{1+\u2}\; \( -i\( u - u^*\), u+u^*, \u2 -1\)
\ee
The matrix $T$ is conjugated to 
$-\tau_3={\rm diag} \( -1,1\)$ as 
\br
T = - W^{\dagger} \; \tau_3 \; W \; ; \qquad \qquad 
W\equiv  \frac{1}{\sqrt{1+\u2}}\; \( 
\begin{array}{cc}
1 & i u\\
i u^* & 1 
\end{array}\) 
\er
and consequently we get the decomposition of the $SU(2)$ group elements
\be
U = W^{\dagger} \; e^{-i \zeta \tau_3}\; W
\ee
Therefore,
\be
U^{\dagger} \pa_{\mu} U =W^{\dagger} \;\( -i\; \pa_{\mu} \zeta \; \tau_3 + 
\frac{2 \sin \zeta}{1+\u2}\(  e^{i\zeta}\; \pa_{\mu} u \; \tau_{+} - 
e^{-i\zeta}\; \pa_{\mu} u^* \; \tau_{-} \) \) W
\ee
with $\tau_{\pm}=\( \tau_1\pm i \tau_2\)/2$.
Denoting $J_{\mu} \equiv W^{\dagger} \;B_{\mu}\; W$, it follows that 
Skyrme's equations of motion are then written as  
\be
D^{\mu} B_{\mu} = \pa^{\mu} B_{\mu} + \sbr{A^{\mu}}{B_{\mu}} = 0 
\lab{zcsky}
\ee
with
\br
A_{\mu} &\equiv & - \pa_{\mu} W\; W^{\dagger} \nonu\\
&=& {1\o{ 1+\mid u \mid^2 }} \(  -i \pa_{\mu} u \, \tau_{+}
-i \pa_{\mu} u^* \, \tau_{-} +  
\h \; \( u \pa_{\mu} u^* - u^* \pa_{\mu} u \) \, \tau_3 \)
\lab{asky}\\
B_{\mu}&\equiv & -i R_{\mu} \tau_3 + \frac{2 \sin \zeta}{1+ \u2}\; \( 
e^{i\zeta}\; S_{\mu} \; \tau_{+} -
e^{-i\zeta}\; S_{\mu}^* \; \tau_{-} \) 
\lab{bsky}
\er
where
\br
R_{\mu}&\equiv& \pa_{\mu} \zeta - 8 \l \; \frac{\sin^2 \zeta}{\(1 + \u2\)^2}
\(  N_{\mu} +  N_{\mu}^*\)\nonu\\
S_{\mu}&\equiv& \pa_{\mu}u + 4 \l \; \( M_{\mu} - \frac{2\sin^2
\zeta}{\(1+\u2\)^2}  \; K_{\mu}\)
\lab{rsdef}
\er
and 
\br
K_{\mu} &\equiv& \( \pa^{\nu} u \pa_{\nu} u^*\) \pa_{\mu} u - \(\pa_{\nu}u\)^2
\pa_{\mu} u^*\nonu\\ 
M_{\mu} &\equiv& \( \pa^{\nu} u \pa_{\nu} \zeta\) \pa_{\mu} \zeta 
- \(\pa_{\nu}\zeta\)^2
\pa_{\mu} u\nonu\\ 
N_{\mu} &\equiv& \( \pa^{\nu} u \pa_{\nu} u^*\) \pa_{\mu} \zeta 
- \(\pa_{\nu}\zeta \pa^{\nu} u\)
\pa_{\mu} u^*
\lab{kmndef}
\er

Notice that $B_{\mu}$ has a vanishing covariant divergence w.r.t. a flat
connection $A_{\mu}$. It has been shown in \ct{afsg} that this
corresponds to a local zero curvature condition  generalizing to higher
dimensions the well known two dimensional Lax-Zakharov-Shabat equation. An
important point of 
the formalism in \ct{afsg} is that 
$B_{\mu}$
should transform under some representation of the algebra defined by
$A_{\mu}$. In fact, the number of conserved currents is equal to the dimension
of the representation. The Skyrme model has three conserved currents and the
representation relevant is the triplet. Indeed, \rf{bsky} lies in the adjoint
of $SU(2)$. There is a systematic algebraic procedure
\ct{erica,gianzo} to find additional conditions for the construction
to hold true in an infinite number of representations. However one can
directly obtain   a submodel of the Skyrme theory with an
infinite 
number of local conserved currents by imposing the following
constraints\footnote{These constraints will suppress the higher spin terms in
the commutator in \rf{zcsky} when extending the representation where $B_{\mu}$
lies, beyond the triplet.}
\be
S_{\mu} \pa^{\mu} u = 0 \; ; \qquad \qquad  R_{\mu} \pa^{\mu} u = 0
\lab{const1}
\ee
and their complex conjugates. Using \rf{kmndef} one notices that 
$K_{\mu} \pa^{\mu} \zeta = N_{\mu}^* \pa^{\mu} u$, $M_{\mu}\pa^{\mu}\zeta =0$,
and $N_{\mu}\pa^{\mu}u =0$\footnote{Other useful relations are:
$K_{\mu}\pa^{\mu}u = 0$, $M_{\mu}\pa^{\mu}u^* + N_{\mu} \pa^{\mu} \zeta =0$}. 
Consequently the second  constraint in \rf{const1}
can also be written as  
\be
S_{\mu} \pa^{\mu} \zeta = 0
\lab{const2}
\ee
By substituting the constraints into \rf{zcsky}, or
equivalently \rf{skyrmeeqom}, one finds the equations of motion for the
submodel to be 
\be
\pa^{\mu} S_{\mu} = 0 \; ; \qquad \qquad 
\pa^{\mu} R_{\mu} = \frac{4 \sin \zeta \cos \zeta}{\( 1+ \u2\)^2}\; 
\( S_{\mu} \pa^{\mu} u^*\) 
\lab{subeq}
\ee
where we have used the fact that 
$S_{\mu} \pa^{\mu} u^* =S_{\mu}^* \pa^{\mu} u$.   

The construction of an infinite number of conserved currents is now quite
simple. Consider the quantity 
\be
\cj_{\mu} \equiv G_1\( \zeta, u,u^*\) S_{\mu} + G_2\( \zeta, u,u^*\) S_{\mu}^*
+ G_3\( u,u^*\) R_{\mu}
\lab{curr}
\ee
with $G_i$ being arbitrary functionals of the fields as shown in \rf{curr},
but not of their derivatives. Using  \rf{const1},  \rf{const2} and  
\rf{subeq} one gets  
\be
\pa^{\mu} \cj_{\mu} = \( \frac{\d G_1}{\d u^*} + 
\frac{\d G_2}{\d u}  + 
\frac{4 \sin \zeta \cos \zeta}{\( 1+ \u2\)^2}\; G_3 \) \( S_{\mu} \pa^{\mu}
u^*\)   
\ee
Therefore, by choosing the functionals such that
\be
\frac{\d G_1}{\d u^*} + 
\frac{\d G_2}{\d u}  + 
\frac{4 \sin \zeta \cos \zeta}{\( 1+ \u2\)^2}\; G_3 = 0
\lab{currcond}
\ee
one gets that $\cj_{\mu}$ is conserved. There are basically two classes of
conserved currents. The first one corresponds to the choice 
$G_1 = \frac{\d G}{\d u}$,  $G_2 = -\frac{\d G}{\d u^*}$ and $G_3 =0$, leading
to the currents
\be
\cj^{G}_{\mu} = \frac{\d G}{\d u} S_{\mu} - \frac{\d G}{\d u^*} S_{\mu}^*
\lab{currg}
\ee
with $G$ being an arbitrary functional of $\zeta$, $u$ and $u^*$, but not of
their derivatives. The second class corresponds to the choice 
$G_a = 4 \sin \zeta \cos \zeta \; H_a \( u,u^*\)$, with $a=1,2$, and  
$G_3 = - \( 1+ \u2\)^2\( \frac{\d H_1}{\d u^*} + 
\frac{\d H_2}{\d u}\)$, leading to the currents
\be
\cj^{(H_1,H_2)}_{\mu} = 4 \sin \zeta \cos \zeta \; \( H_1 S_{\mu} + H_2
S^*_{\mu} \) - \( 1+ \u2\)^2\( \frac{\d H_1}{\d u^*} + 
\frac{\d H_2}{\d u}\) \; R_{\mu}
\lab{currh1h2}
\ee
with $H_a$ being arbitrary functionals of $u$ and $u^*$, but not of their
derivatives. The three Noether currents $J_{\mu}$ in \rf{skyrmeeqom} are
linear combinations of currents belonging to the two classes \rf{currg} and
\rf{currh1h2}.   

In terms of the fields, the first constraint in \rf{const1}, namely $S_{\mu}
\pa^{\mu} u = 0$, reads   
\be
\( \pa_{\mu} u\)^2 = -4 \l \; \frac{\( \pa_{\mu}u \pa^{\mu} \zeta\)^2}{1-4\l
\(\pa_{\nu} \zeta\)^2 }
\lab{const1a}
\ee
and the second constraint in \rf{const1}, after $\( \pa_{\mu} u\)^2$ is
eliminated using \rf{const1a}, reads 
\be
\( \pa_{\mu}u \pa^{\mu} \zeta\)\( 1 - 8 \l \frac{\sin^2 \zeta}{\(1+\u2\)^2} 
\( \pa_{\nu} u \pa^{\nu} u^* + 4 \l \frac{\( \pa_{\nu}u \pa^{\nu} \zeta\)\(
\pa_{\rho}u^* \pa^{\rho} \zeta\)}{1-4\l\(\pa_{\s} \zeta\)^2 }\)\) =0
\lab{const1b}
\ee
Therefore, the conditions 
\be
\( \pa^{\mu} u \)^2 = 0 \; ; \qquad \qquad \pa^{\mu} \zeta \; \pa_{\mu} u = 0
\lab{sufcond}
\ee
are sufficient for the constraints \rf{const1} to be satisfied. 
In the static case however, the conditions \rf{sufcond} are equivalent to the
constraints \rf{const1a} and \rf{const1b}. Indeed, one can easily show, using
the real part of \rf{const1a}, that
for fields not depending upon time 
\br
1 &-& 8 \l \frac{\sin^2 \zeta}{\(1+\u2\)^2} 
\( \pa_{\nu} u \pa^{\nu} u^* + 4 \l \frac{\( \pa_{\nu}u \pa^{\nu} \zeta\)\(
\pa_{\rho}u^* \pa^{\rho} \zeta\)}{1-4\l\(\pa_{\s} \zeta\)^2 }\) = \nonu\\
1 &+& 16 \l \frac{\sin^2 \zeta \( \vecgrad u_1\)^2}{\( 1+\u2\)^2}\; 
\( 1 - \frac{\(\vecgrad u_1 \cdot \vecgrad
\zeta\)^2}{\(\vecgrad u_1\)^2\(\vecgrad \zeta\)^2 } \; 
\frac{4\l \(\vecgrad \zeta\)^2}{1 + 4\l \(\vecgrad \zeta\)^2}\) \geq 1
\er
where we have denoted $u=u_1 + i u_2$, 
and $\vecgrad f$ denotes the spatial gradient of $f$. Consequently, the
constraints \rf{const1a} and \rf{const1b} can only be satisfied if 
\be
\vecgrad u \cdot \vecgrad u =0 \; ; \qquad \qquad 
\vecgrad u \cdot \vecgrad \zeta =0
\lab{staticconst}
\ee

Notice that the conditions \rf{staticconst} can be easily solved by an Ansatz
which  
contains as a particular case the rational map Ansatz used to approximate the
soliton solutions \ct{manton,sut}.  We denote the three Cartesian space
coordinates as $x_i$, $i=1,2,3$, and the radial variable as $r\equiv
\sqrt{x_1^2 + x_2^2 + x_3^2}$. We shall use the stereographic projection to
parametrize the points on a sphere of radius $r$ by a complex coordinate
$z$. So we define
\be
\frac{{\vec x}}{r} \equiv 
\frac{1}{1+\z2}\; \( -i\( z - z^*\), z+z^*, \z2 -1\)
\ee
The metric is then given by
\be
ds^2 = \( d r\)^2 + \frac{4 r^2}{\( 1+\z2\)^2}\; dz \; dz^*
\lab{metric}
\ee
We consider time independent solutions such that 
\be
\zeta = \zeta \( r\) \qquad u = u \( z \) \qquad u^* = u^* \( z^* \)
\lab{ansatz}
\ee
Then due to the metric \rf{metric}, it follows that \rf{staticconst},  and
consequently the constraints \rf{const1}, are satisfied. 

Let us now analyze the configurations \rf{ansatz} that satisfy the equations
of motion 
\rf{subeq} of the submodel. Substituting \rf{ansatz} into \rf{rsdef} one finds
that
\br
S_0= S_r=S_{z^*}=0 \; ; \qquad 
S_{z}= \( 1 + 4 \l \( \(\pa_r\zeta\)^2  + 
\frac{\sin^2 \zeta}{r^2} \; \b   \)\)
\; \pa_{z} u
\er
where
\be
\b \equiv \frac{\(1+\z2\)^2}{\(1+\u2\)^2}\; \pa_{z} u  \pa_{z^*} u^* 
\lab{betadef}
\ee
Then the  first equation in \rf{subeq} and its complex conjugate imply that
\be
\pa_z \, \b = \pa_{z^*} \, \b = 0
\ee
Since by definition, $\b$ is not a function of $r$, it follows that $\b = {\rm
constant}$. The second equation in  \rf{subeq} gives the equation for the
profile function $\zeta$ as
\be
\( 1+ 8 \l \b \frac{\sin^2 \zeta}{r^2}\) \zeta^{\pr\pr} + 
4 \l \b \frac{\sin 2 \zeta}{r^2} \(\zeta^{\pr}\)^2 + \frac{2}{r}\zeta^{\pr}
- \b \frac{\sin 2\zeta}{r^2} - 
4 \l \b^2 \frac{\sin 2 \zeta \sin^2\zeta}{r^4} =0
\lab{profile}
\ee
Notice that due to \rf{ansatz} and \rf{betadef} we can write 
\be
\frac{2i dz^*\wedge dz}{\( 1 +\z2\)^2} \, \b = 
 \frac{2i du^*\wedge du}{\( 1 +\u2\)^2}
\ee
and so it follows that $\b$ is the pull-back of the
area form on the target sphere of the map $z\ra u$. Therefore, the topological
charge becomes
\be
Q = \frac{1}{4 \pi^2} \, \int 
\frac{2i du^*\wedge du\wedge d\zeta}{\( 1 +\u2\)^2} = 
\frac{\b}{4 \pi^2} \, \int \frac{2i dz^* \wedge dz\wedge dr}{\( 1 +\z2\)^2}
 \, \zeta^{\pr} = \b \; \frac{\zeta\(0\) - \zeta\( \infty \)}{\pi}
\ee

For the hedgehog solution  $u=z$, and so  $\beta=1$, the equation \rf{profile}
becomes the well known ODE for the original Skyrme profile 
function. With the standard boundary conditions 
 $\zeta(\infty)=0$ and  $\zeta(0)=\pi$, one then gets $Q=1$. 
In order to get the solution with $Q=-1$ one has to interchange $z$
by $z^*$ in \rf{ansatz} and take $u=z^*$. Consequently, the Skyrmions with
charges $Q=\pm 1$ are solutions of our submodel. 

The question now is if there are other holomorphic  solutions  within the
Ansatz \rf{ansatz}. The answer is that only those configurations given by
\be
u = \frac{a z + b}{c z + d} \qquad \qquad ad - bc = 1
\lab{mobius}
\ee
lead to solutions, and therefore one obtains through \rf{ansatz} solutions of
charge $Q=\pm 1$ only. Notice that \rf{mobius} correspond to rotations of the
solution $u=z$.  
The simplest way to prove it is to observe that the
map $z\ra u$ provided by the configuration implies that the metric on the two
spheres are related by
\be
\frac{du\, du^*}{\( 1 + \u2\)^2} = \b \;
\frac{dz\, dz^*}{\( 1 + \mid z\mid^2\)^2}
\ee
But the only maps satisfying that are the modular transformations \rf{mobius}
which imply that $\b =1$. 
Notice that the rational map Ansatz $u =\frac{p\(
z\)}{q\(z\)}$, with $p$ and $q$ being
 polinomials in $z$ with degree higher than
one, are particular cases of \rf{ansatz}. Consequently they can not lead to
exact solutions of the Skyrme model, even though  they constitute good 
approximations to the minimal energy numerical configurations \ct{sut,manton}.
It would be  very interesting to 
check if the numerical higher charge solutions in \ct{sut} satisfy the 
static constraints 
\rf{staticconst}, which has a much simpler form than the full constraints
\rf{const1} since they do not involve the scale determined by the coupling
constant $\l$.  A positive outcome of such check would encourage the use of 
those constraints in the construction of new Ans\"{a}tze for the higher 
charge solutions.   

In conclusion, we have identified an integrable submodel containing the
Skyrmions of charges $Q=\pm 1$, showing that they are the only solutions within
the meromorphic Ansatz \rf{ansatz}. The submodel, which  is   
defined by the 
constraints \rf{const1} and equations of motion \rf{subeq},   possesses the
infinite set of local conserved 
currents given by \rf{currg}-\rf{currh1h2}. That is our  main result since it
unravels an infinite symmetry inside the Skyrme model which  opens  ways to
the construction of exact methods.

\lskip
{\bf Acknowledgments.} JSG gratefully acknowledges Fapesp (Proj. Tem\'atico
98/16315-9) for financial
support and IFT/UNESP for hospitality. LAF is very grateful to the Dept. of
Physics of Univ. Santiago de Compostela for the hospitality and acknowledges
the financial support through the Spanish CICYT AEN99-0589 grant. We are 
grateful to O. Alvarez and P.M. Sutcliffe for discussions.

\lskip

\end{document}